\documentclass{iaus}
\usepackage{graphicx}

\title[Observations of magnetic fields in hot stars] 
{Observations of magnetic fields in hot stars}

\author[Petit, V.]   
{V. Petit$^1$
 }

\affiliation{$^1$Dept. of Geology \& Astronomy, West Chester University, West Chester, PA 19383, USA  \\ email: {\tt VPetit@wcupa.edu} \\[\affilskip]
}

\pubyear{2010}
\volume{272}  
\pagerange{119--126}
\jname{Active OB stars: structure, evolution, mass loss, and critical limits}
\editors{C. Neiner, G. Wade, G. Meynet \& G. Peters, eds.}
\begin{document}

\maketitle

\begin{abstract}
The presence of magnetic fields at the surfaces of many massive stars has been suspected for decades, to explain the observed properties and activity of OB stars. However, very few genuine high-mass stars had been identified as magnetic before the advent of a new generation of powerful spectropolarimeters that has resulted in a rapid burst of precise information about the magnetic properties of massive stars. During this talk, I will briefly review modern methods used to diagnose magnetic fields of higher-mass stars, and summarize our current understanding of the magnetic properties of OB stars. 
\end{abstract}

\firstsection 

\section{Introduction}

Hot, massive stars have an enormous impact on their galactic environment. Energy and momentum are injected into the surrounding interstellar medium by their powerful stellar winds and radiation fields during their short, but nonetheless colourful, lives and by the dramatic core-collapse supernovae that mark their deaths. In this way, they seed the interstellar medium with the products of their nucleosynthesis, to be recycled into the next generations of stars and planets, hence playing a key role in the chemical enrichment of the Universe. These rapidly-evolving stars thereby drive the chemistry, structure and evolution of galaxies, dominating the ecology of the Universe - not only as supernovae, but also during their entire lifetimes - with far-reaching consequences.

The evolution of a massive star is strongly determined by its rotation, as well as the mass lost through its stellar wind, both of which can be influenced by the presence of a magnetic field. A field can couple different layers of a star's interior, hence modifying internal differential rotation (Maeder and Meynet 2005). If a field has a large-scale component that extends outside the stellar surface, it can also channel a stellar wind, creating a structured wind - a magnetosphere - which will modify the rate and geometry of mass loss. Furthermore, if the field couples the rotating surface of the star with its outflowing stellar wind, both effects will result in a different angular momentum loss (via the outflowing stellar wind) than that of a non-magnetic star (ud-Doula et al. 2009). As angular momentum and mass loss are determing factors in stellar evolution calculations, it is crucial that the effect of magnetic field be understood properly in order to correctly use evolutionary tracks and isochrones when interpreting, for example, large datasets of OB associations (see Evans this proceeding).

In the last decade, our knowledge of the basic statistical properties of massive star magnetic fields has significantly improved, in part due to a new generation of powerful spectropolarimetric instrumentation. 
In this paper I will review modern methods used to diagnose magnetic fields of higher-mass stars, and briefly summarize our current understanding of the magnetic properties of OB stars. 

\section{Zeeman effect}

The best way to directly detect a stellar magnetic field is by the Zeeman effect. 
When light passes through a medium and forms a spectral line by a atomic transition, the radiative transfer will be modified by the presence of the field, which split the energy levels in multiple components.

The first effect will be a splitting of the spectral line in multiple component. The width of the splitting $\delta\lambda_z$ is proportional to the modulus of the field:
\begin{equation}
 \Delta\lambda_z \propto \lambda_0 \bar{g} |\vec{B}|,
\end{equation}
where $\lambda_0$ is the wavelength of the unperturbed transition and $\bar{g}$ is the effective Land\'e factor of the transition. 
The splitting is therefore a scalar quantity that is not affected by the orientation of the field. However, the typical spectral separation will only be about 1-2\,km\,s$^{-1}$ per kilogauss in the optical domain. So, unless the field is quite strong, the splitting will be less than the typical Doppler broadening of the line profiles of a hot star. 

\begin{figure}[t]
\begin{center}
 \includegraphics[width=4.0in]{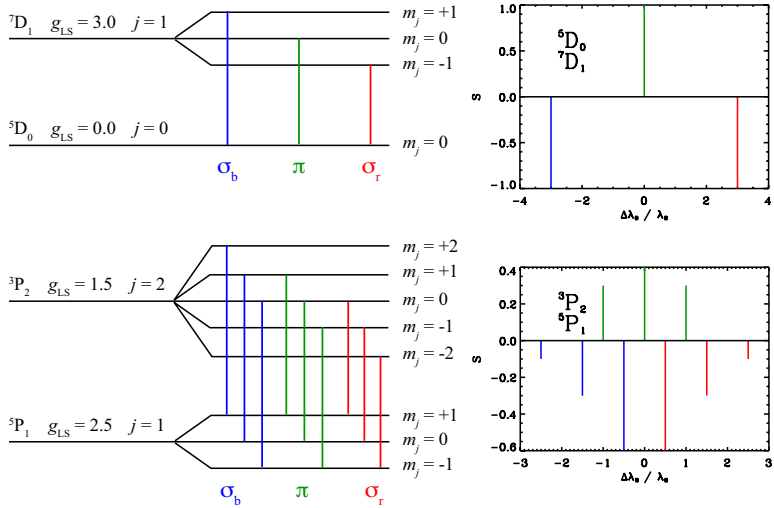}  
 \caption{\textit{Left:}  Zeeman splitting level separation and admissible sub-transitions for the transitions $^7$D$_1$\,-\,$^5$D$_0$ and $^3$P$_2$\,-\,$^5$P$_1$. The sub-transitions that correspond to $\Delta m_j=-1, 0, +1$ are named $\sigma_r$, $\pi$ and $\sigma_b$, respectively. The Land\'e factors of each level are computed under $LS$ coupling. 
   \textit{Right:} Zeeman patterns corresponding to the illustrated transitions. The component separations are expressed in terms of Lorentz units ($\lambda_B$). By convention, the $\pi$ components are illustrated upward, and the $\sigma$ components downward.
	 } 
\label{fig_zeeman}
\end{center}
\end{figure}

In addition to the line splitting, the multiple Zeeman components of the transition, illustrated in Fig. \ref{fig_zeeman} will have different polarisation states. The effect of the field can be decomposed in two contributions. The component of the field that is along the line of sight, the longitudinal field $B_{||}$, will partially circularly polarise each of the $\sigma$ components. The polarisation of the $\sigma_b$ component will be orthogonal to that of the $\sigma_r$ component. As the Stokes $V$ parameter is the subtraction of the two orthogonal states, we will observe a net change in Stokes $V$ across the line profile. It is important to keep in mind that the circular polarisation produced is not only dependent on the field strength, but also on the orientation of the field with respect to the observer.

The component of the field that is perpendicular to the line of sight, the transverse field $B_\perp$, will produce linear polarisation along the transverse field axis for the $\sigma$ components and perpendicular to the field axis for the $\pi$ components. There will therefore be a change in the Stokes parameters $Q$ and $U$ across the line profile. The amount of polarisation in each linear Stokes parameters depends on the orientation of the transverse field on the plane of the sky. It is important to note than for the same field strength in the longitudinal and transverse directions, the linear polarisation produced will be substantially weaker than the circular polarisation.

Circular polarisation observations are therefore a really good choice in order to detect magnetic fields. Furthermore, given the accessibility of the four Stokes parameters ($I$, $V$, $Q$ and $U$), it is in principle possible to characterize completely a magnetic field's strength and orientation.
However, it is not as simple as it looks. It is important at this point to remember that the light coming from a spatially unresolved star is in fact the combination of the light produced at each point on the visible stellar surface - which can posses different emission properties, as well as different local field strengths and orientations.

As the polarisation produced by a certain point on the stellar surface is sensitive to the orientation of the local field vector at that point, some cancellation can occur in the disc-veraged profile with another point with opposite field orientation, hence making the global observed polarisation more difficult to detect and more difficult to interpret.

\begin{figure}[t]
\begin{center}
 \includegraphics[width=4.0in]{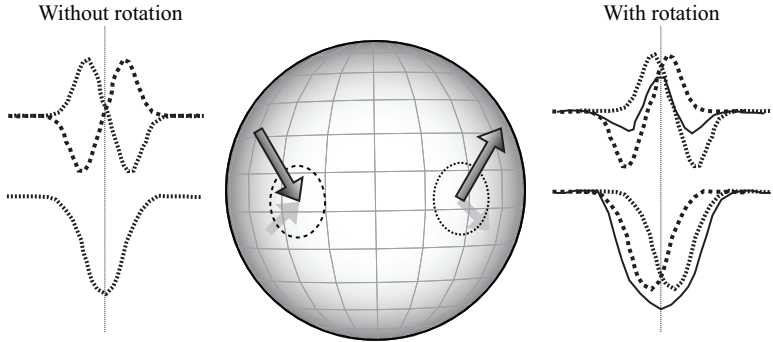}  
 \caption{Illustration of the effect of rotation on local line profiles. In the case without rotation (left), two local field vectors with opposite orientations (indicated by the arrows) will produce overlapping Stokes $I$ profiles. The local Stokes $V$ profiles will cancel out (dotted and dashed lines). When the star is rotating the local profiles will not overlap because of the induced Doppler shifts The Stokes $V$ profiles will not cancel out perfectly, leaving some net circular polarisation (solid line).   }
\label{fig_cross}
\end{center}
\end{figure}

The rotation of the star is also important to consider. If the star is not seen pole-on, the line-of-sight components of the rotation velocity will introduce a Doppler shift distribution across the stellar disk. 
Hence, the contribution to the polarisation coming from two regions of opposite local field orientation might not cancel out perfectly, as shown in Fig. \ref{fig_cross}. Therefore, a polarisation signal across the line profile may  still be seen even if the net magnetic field over the stellar disk in the line-of-sight direction is null. The required line-of-sight projected rotation is only of a few km\,s$^{-1}$.

For rotating stars, the morphology of the Stokes $V$ profile therefore contains information about the distribution of magnetic field strength and orientation over the visible stellar disc. Our ability to access this information depends not only on the structure of the field and the rotation of the star, but also the resolving power of the instrument used to observe it.
In practice, we need to divide the spectropolarimetric instruments in two categories, depending on their capacity to resolve the line profile in the spectral domain. 
If the resolution elements of the instrument are wider than the Doppler shifts introduced by rotation, all the spatial information is lost. Therefore, lower resolution instrument are only sensitive to the net longitudinal field component across the stellar disk.  
Table \ref{tab_ins} lists the spectropolarimetric instruments currently in use, as well as the resolution and the width of their resolution elements. 

\begin{table}
  \begin{center}
  \caption{Non-exhaustive list of spectropolarimetric instruments.}
  \label{tab_ins}
 {\scriptsize
  \begin{tabular}{ l l c c }\hline 
\textbf{Instrument} & \textbf{Telescope} & \textbf{Resolving power} & \textbf{Resolution element (km\,s$^{-1}$)} \\
  \hline
FORS1 \& FORS2 $^1$ & VLT (8.2\,m) & $<2\,000$ & 150 \\
ISIS & WHT (4.2\,m) & $< 10\,000$ & 30 \\
Spectropolarimeter & DAO/Plaskett (1.8\,m) & $< 10\,000$ & 30 \\
ESPaDOnS & CFHT (3.6\,m) & 65\,000 & 5 \\
Narval & TBL (2\.m) & 65\,000 & 5 \\
SEMPOL/UCLES & AAT (3.9\,m) & 70\,000 & 4 \\
HARSPol & ESO-3.6 (3.6\,m) & 70\,000 & 4 \\
NES & SAO/BTA (6\,m) & 60\,000 & 5 
 \\ \hline
  \end{tabular}
  }
 \end{center}
\vspace{1mm}
 \scriptsize{
 \textit{Notes:}\\
  $^1$ FORS1 has been decommissioned, and replaced by FORS2, which has similar characteristics. \\
  }
\end{table}

\section{Low-resolution instruments}

\begin{figure}[t]
\begin{center}
 \includegraphics[width=5in]{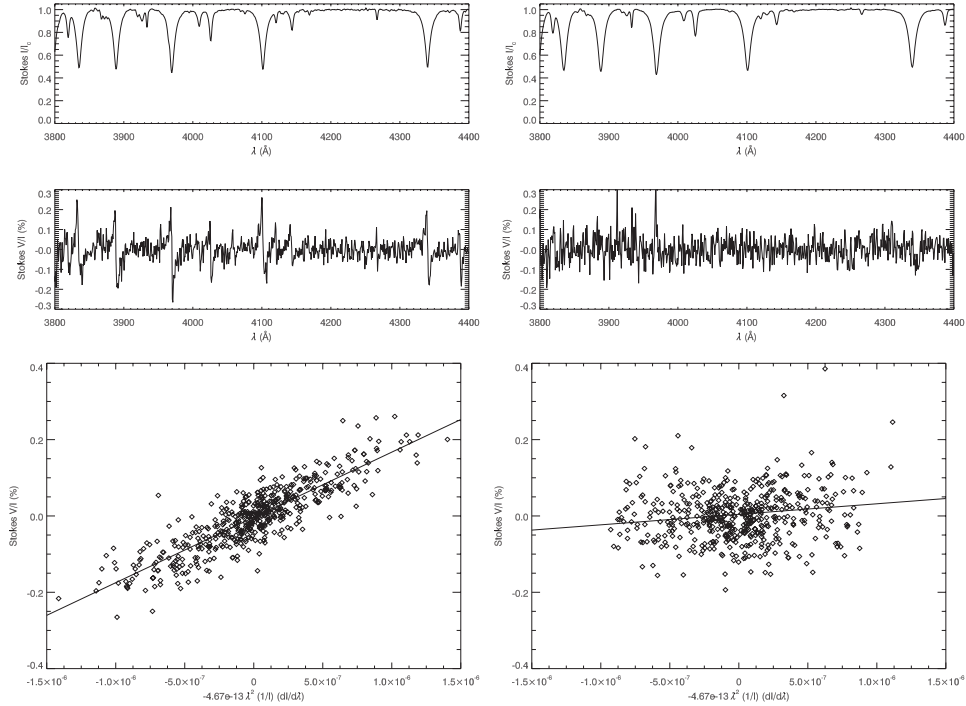}  
 \caption{FORS1 observations of the B-type stars NGC\,3766~170 (left) and NGC\,3766~94 (right).  \textit{Top:} Stokes $I/Ic$ spectrum. \textit{Middle:} Stokes $V/Ic$ spectrum. \textit{Bottom:} The global longitudinal field $B_l$ is proportional to the slope of the least-squares linear fit to the observed data. The results are $B_l=1\,710\pm32$\,G (53$\sigma$) and $B_l=276\pm55$\,G (5$\sigma$), respectively. From McSwain (2008)   }
\label{fig_fors}
\end{center}
\end{figure}

Fig. \ref{fig_fors} (left) shows an example of a FORS1 spectrum (low-resolution) of a strongly magnetic Bp star (kG level), where changes of the circular polarisation across the line profiles can be clearly seen. 
In order to interpret this polarisation in term of a interesting magnetic quantity, the method described by Bagnulo et al. (2002) is generally used.  The idea is to determine the longitudinal component value that would provide such a Stokes $V$ profile if it was \textit{local}, in the weak field approximation (then the splitting is negligible). It has been shown (see Landstreet 1982) that this approximation will result in a line-strength weighted mean of the longitudinal field integrated over the stellar disc. This value is usually simply referred to as the global longitudinal field $B_l$ in the literature.  In order to interpret this global longitudinal field in terms of the magnetic field at the stellar surface, one must then make hypothesis about the structure of the field and also about the emission properties of the stellar disc. For example, a 50 G measurement of the global longitudinal field could reflect a surface magnetic field of relatively simple topology (i.e. relatively little global cancellation) with a strength of order 100 G, or alternatively a rather highly structured field (i.e. lots of global cancellation) with a much greater strength.

As it does not make assumptions about the surface topology of the field itself, the global longitudinal field value is a really useful value to find rotational periods, as the visible field changes as the star rotates (e.g. Bychkov et al. 2005) . It is also possible to perform some modelling, by assuming some geometry (e.g. Landstreet \& Mathys 2000). This value is also useful as a basis for statistical studies (e.g. Landstreet et al. 2007, Kholtygin et al. this proceeding). 

The observed circular polarisation $V/I$ is related to the global longitudinal field by:
\begin{equation}
	\frac{V}{I} = -\bar{g}\, C_z\, \lambda^2\, \frac{1}{I}\, \frac{\mathrm{d}I}{\mathrm{d}\lambda}\,B_l,~~~~
	C_z= \frac{e}{4\pi m_ec^2}~(\simeq4.67\times10^{-13}\mathrm{\AA}^{-1}),
\end{equation}
where $e$ is the electron charge, $m_e$ the electron mass and $c$ the speed of light (Bagnulo et al. 2002). 
The strategy is to take each point in the spectrum and plot the value of $V/I$ versus $(\mathrm{d}I/\mathrm{d}\lambda)/I$ and solve for the slope, which yields a value for $B_l$ (see Fig. \ref{fig_fors} bottom left).

In some cases, the magnetic field present is quite evident, just by a quick glance at the polarisation spectrum. However, as the we are trying to detect magnetic fields at the limit of the instrument's capacity, we need to rely on statistics (for example, see Fig. \ref{fig_fors}, right panels). It is customary to use the derived global longitudinal field value as a detection diagnostic, by looking at the statistical significance of the field detection. 
Therefore, a realistic and complete treatment of the uncertainties is required. Novel methods that have been put forward recently (e.g. Rivinius et al. 2010) have shown that $B_l$ error bar of FORS1 data may have been significantly underestimated by some authors (see Rivinius et al. this proceeding). Re-analysis of archived FORS1 observations has identified spurious detections in the literature (Bagnulo et al. in prep, Fullerton et al. this proceeding). This could explain reported magnetic field detections that have not been confirmed by high-resolution observations (see Silvester et al. 2009).

\section{High-resolution instruments}

\begin{figure}[t]
\begin{center}
 \includegraphics[width=4.5in]{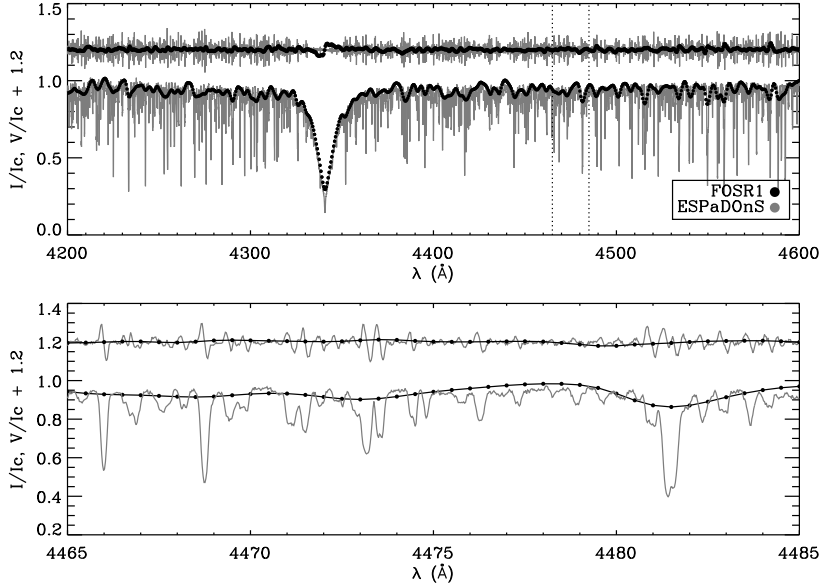}  
 \caption{FORS1 and ESPaDOnS observations of HD\,94660. \textit{Top: } portion of the Stokes $I$ and Stokes $V$ (shifted vertically by 1.2) spectra, observed by FORS1 at VLT (in black) and by ESPaDOnS (in grey). \textit{Bottom: } enlargement of a part of the spectrum (as shown by dotted lines on the top panel) to illustrate the higher resolution of ESPaDOnS. Note the multitude of spectral lines, and their associated Stokes $V$ signatures, that are clearly evident in the high-resolution spectrum, but that blend together almost of invisibility in the low-resolution spectrum.}
\label{fig_lowres}
\end{center}
\end{figure}

Fig. \ref{fig_lowres} shows a comparison of two spectra of the magnetic Bp star HD\,94660, obtained with the low resolution instrument FORS1 at VLT (in black) and the high-resolution intrument ESPaDOnS at CFHT (in grey). 
With the higher resolution, the multitude of metallic lines can be resolved. Information about the velocity of features in the line profile can therefore be translated into a spatial location on the stellar disk. One also has the ability to clearly identify which lines are present in the spectrum, and to diagnose the field using specific, clearly resolved features. 

\begin{figure}[t]
\begin{center}
 \includegraphics[width=4.75in]{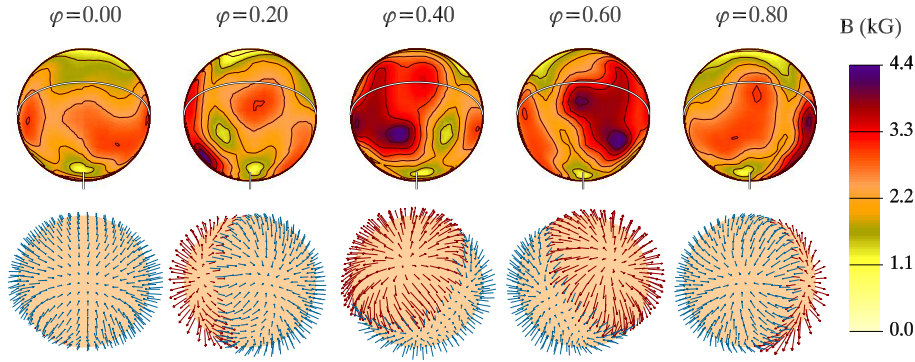}  
 \caption{Surface magnetic field distribution of the Ap star $\alpha^2$\,CVn derived from an high-resolucion MuSiCoS time-serie of all four Stokes parameters. The star is shown at the five rotational phases indicated at the top of the figure. The upper row represent the distribution of field strength, with contour of equal magnetic field strength plotted every 0.5\,kG. The lower panel shows the orientation of the magnetic vectors. From Kochukhov \& Wade (2010).}
\label{fig_oleg}
\end{center}
\end{figure}

With a rotation resolved time-series of high-resolution spectra in circular or circular and linear polarisation (i.e. Stokes $IV$ or Stokes $IVQU$), it is possible to perform a magnetic modelling that solves the radiative transfer problem simultaneously for for detailed distribution of both magnetic field and other relevant quantities (e.g. chemical abundance) over a star's surface (Piskunov \& Kochukhov 2002). The result of such an analysis is a map of abundance in diverse chemical elements as well as a map of the surface magnetic field (e.g. Kochukhov et al. this proceeding). Fig \ref{fig_oleg} show such a map of the surface field of the Ap star $\alpha^2$\,CVn (Kochukhov \& Wade 2010).

Because the polarisation induced in spectral lines by the Zeeman effect is relatively weak (less than 0.1\$ for a 1\,kG field for a strong metallic spectral of a star rotating with a $v\sin i$ of 50\,km\,s$^{-1}$), it is not always possible it is not always possible to measure the polarisation induced in individual spectral lines. As most of the high-resolution spectropolarimeters currently available are echelle spectrometers, they cover a large spectral domain. It is then possible to increase the magnetic sensitivity by the simultaneous use of many lines present in a spectrum.
Currently, the most widely-used and powerful technique is the Least-Squares Deconvolution (LSD) procedure, as described in Donati et al. (1997). 
The main assumption of the LSD method is that the shape of the line profile in intensity and circular polarisation is roughly the same for all the lines, and that this shape is scaled by a weight in order to give the observed profiles. These weights are proportional to the line depth for Stokes $I$ and proportional to the product of the line depth, the Land\'e factor and the wavelength for Stokes $V$. 
From a list of spectral lines, a small window is cut out of the observed Stokes $I$ and $V$ spectrum and is converted in velocity space for each spectral line. One could imagine that all these windows are then averaged together, in order to produce a mean, high s/n profile. In practice, the shape that best fits the ensemble of single line windows is extracted analytically by the least-squares method.

The presence of a signal in the circular polarisation LSD profile is usually diagnosed by looking at the deviation of the signal from zero. This is quantified by the probability that a deviation as large as the one encountered can be produced by random noise. Generally, a signal is considered definitively detected when this probability gets lower than $1\times10^{-5}$ (i.e. 0.001\%).

It is possible to obtain a value equivalent to the global longitudinal field measured by low-resolution instruments by calculating the first moment of the Stokes $V$ profile, as described by Donati et al. (1997) and modified by Wade et al. (2000):
\begin{equation}
B_l = -2.14\times10^{11} \frac{ \int vV(v)\mathrm{d}v}{ \bar{g}\, \lambda\, c\, \int (1-I(v))\mathrm{d}v}~~[\mathrm{G}].
\end{equation}
 As high-resolution instruments can detect net circular polarisation across the line profile even when the global longitudinal field is null, it is important to note that the detection of the Stokes $V$ signature provides a more robust field detection diagnostic - a qualitatively new diagnostic that is only available from high-resolution data.

The question is now, what is the meaning of the derived LSD shape?
The general assumption as been that the LSD profile is equivalent to a real spectral line. It has been shown however that for hot stars this assumption is crude and gives satisfactory results only in a few circumstances, for example at field strengths lower than $\sim2$\,kG when considering Stokes $V$ (Kochukhov et al. 2010).
Some extra care must therefore be taken when interpreting LSD profiles to derive magnetic properties. 

\begin{figure}[t]
\begin{center}
 \includegraphics[width=2.5in]{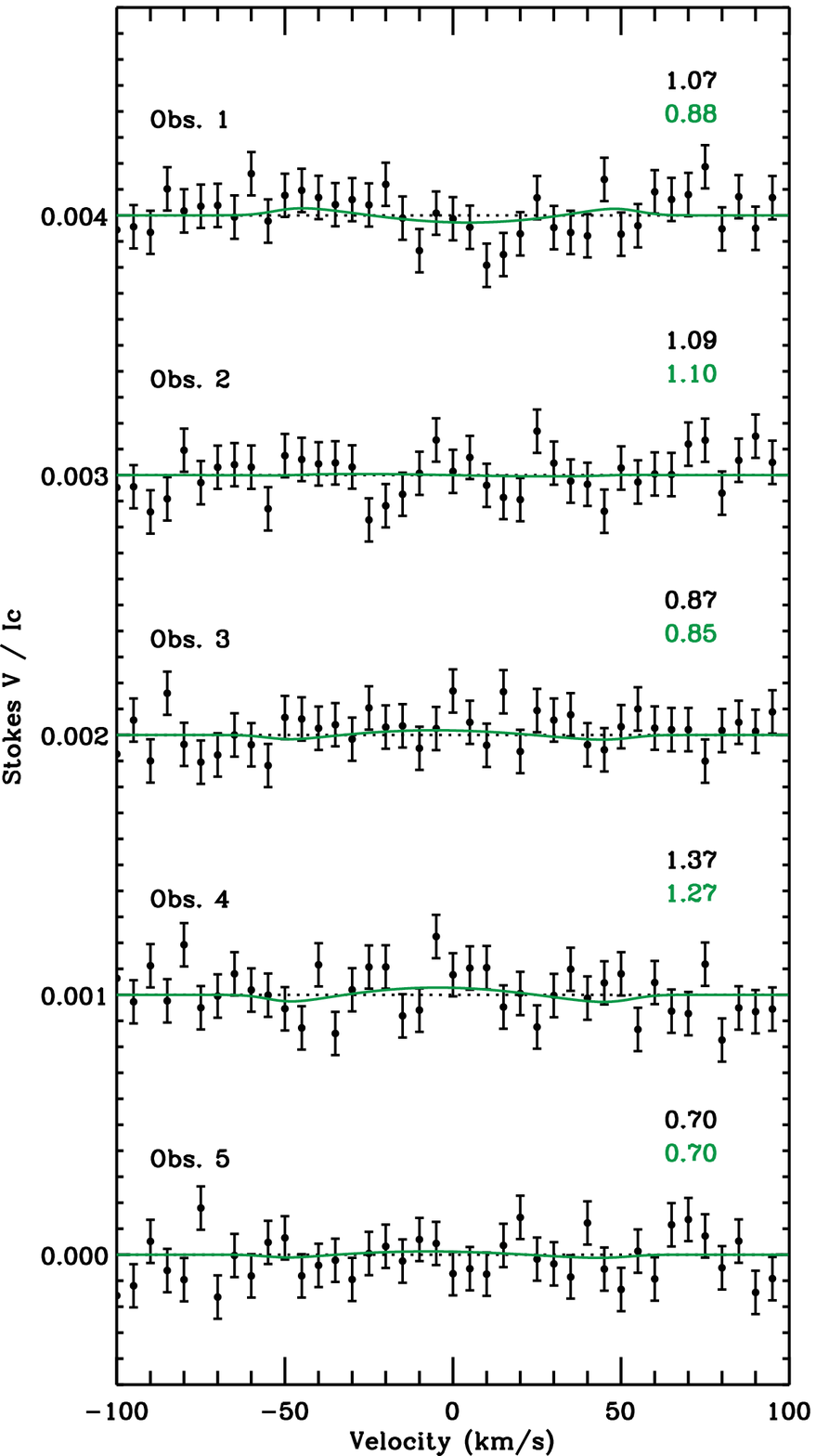}  
 \includegraphics[width=2.5in]{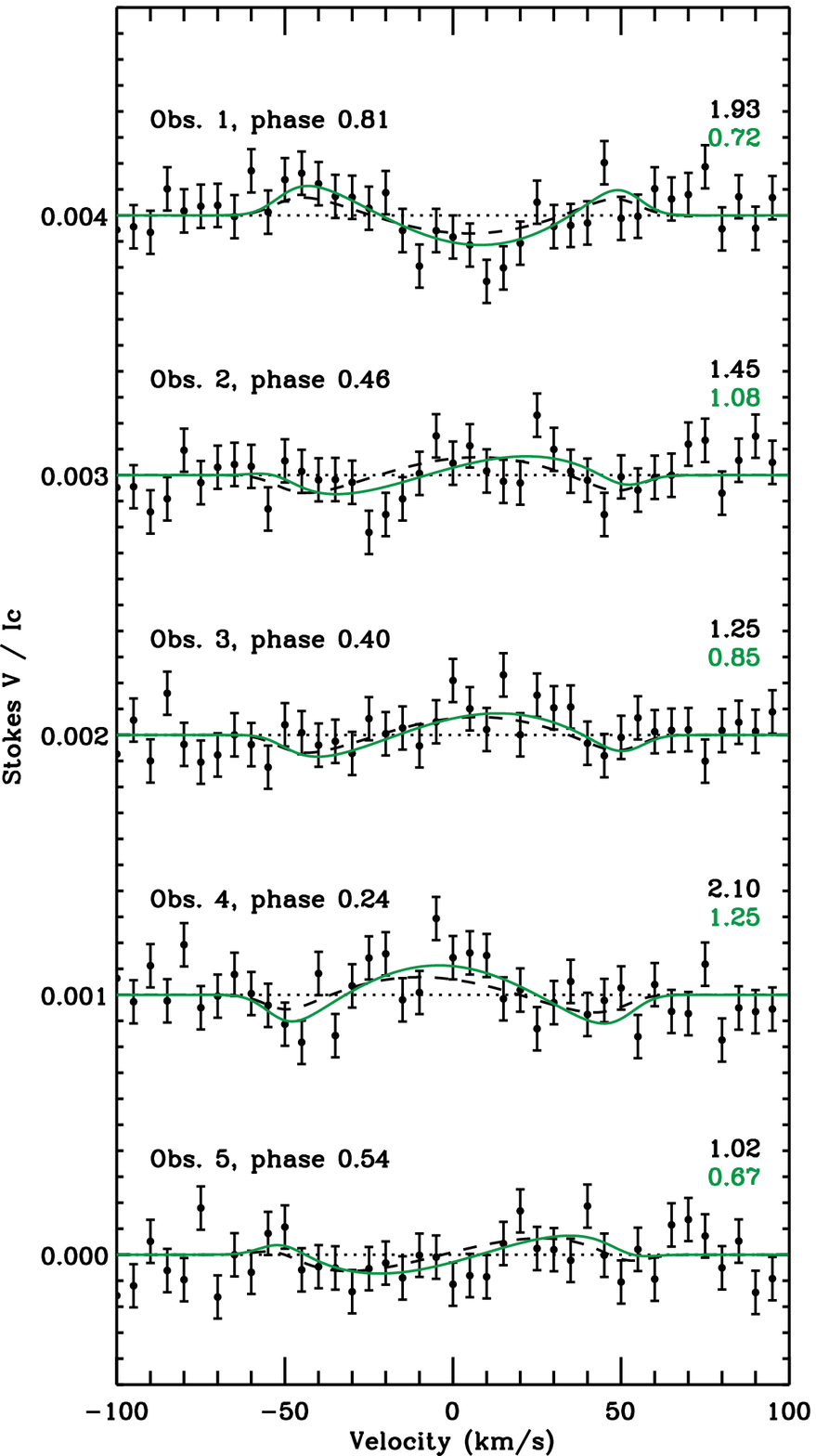}
 \caption{ Simulated Stokes $V$ data for a non-magnetic star (left) and for a dipolar field of 125\,G for 5 randomly chosen rotational phases (right). The underlying magnetic profile is shown in dashed lines. The dotted line shows the no magnetic field model ($M_0$) and its associated reduced $\chi^2$ is indicated at the top. The full line show the best dipole configuration (model $M_1$) for the combined observations and its $\chi^2$ is indicated at the bottom. The odds ratio for the non-magnetic case is $\log(M_0/M_1)=0.295$. The odds ratio for the magnetic case is $\log(M_0/M_1)=-9.05$ (i.e. the magnetic model is 9 orders of magnitude more preferred in the second case). The underlying magnetic signal can therefore be detected, even if none of the individual observation leads to a formal detection. From Petit et al. in prep. }
\label{fig_bayes}
\end{center}
\end{figure}

Nevertheless, LSD is a powerful technique to detect weak magnetic fields in hot stars (e.g. Henrichs et al this proceeding; Grunhut et al. this proceeding; Petit et al. this proceeding), and also to some extent characterize the surface field with the means of phase-resolved Stokes $V$ time series (for example, see the complex surface field of $\tau$\,Sco by Donati et al. 2006b)
We can also add some information to the formal $\chi^2$ statistics used for detection, by 
using knowledge of the shape of an expected deviation in Stokes $V$. With multiple noisy observations, it is possible to pick up an underlying signal by computing the odd ratios of the no magnetic field model ($M_0$) to the inclined dipole model ($M_1$), in a Bayesian framework (Fig. \ref{fig_bayes}). 
As the exact rotation phases of stars are not generally known in advance, observation needs to be compared with the observed Stokes $V$ profiles to a rotation independent geometry (see Petit et al. 2008 for the dipolar oblique rotator case).
Furthermore, by performing a Bayesian parameter estimation for the dipole model, it is possible to obtain an estimate of the dipole field strengths admissible by observations. This is useful for preliminary analysis of sparse dataset (e.g. Grunhut et al. 2009), or to derive upper limits for a non-detection (e.g. Fullerton et al. this proceeding; Petit et al. this proceeding; Shultz et al. this proceeding)

\section{A family portrait of stellar magnetism}

In the sun and essentially all other cool, low-mass stars, vigorous magnetic activity results from 
the conversion of convective and rotational mechanical energy into magnetic energy, generating 
highly structured and variable magnetic fields whose properties correlate strongly with stellar mass, age and rotation rate (e.g. Hartmann \& Noyes 1987). Although the dynamo mechanism that drives this process is not understood in detail, its basic principles are well established. 

The magnetic fields of higher-mass stars (above about 2 solar masses, in which the energy flux 
from the convective envelope begins to vanish, and in which no fully-convective pre-main sequence Hayashi phase is experienced) are qualitatively different from those of cool, low-mass stars. They are detected in only a small fraction of stars, they are structurally much simpler, and frequently much stronger, than the fields of cool stars. Most remarkably, their characteristics show no clear correlation with basic stellar properties,in particular mass and rotation rate. (e.g. Kochukhov \& Bagnulo 2006; Landstreet et al. 2007).  

The weight of opinion currently holds that these puzzling characteristics reflect a fundamentally 
different field origin than that of cool stars: that the observed fields are not generated by dynamos, but rather that they are fossil fields; i.e. remnants of field accumulated or generated during star formation (e.g. Braithwaite 2009; Duez \& Mathis 2010). Although the fossil paradigm provides a useful framework for interpreting the large-scale magnetic fields of higher-mass stars, other dymano related process could also be at work on smaller scales, or within the convective core of the star (see Mathis et al. this proceeding; Cantiello this proceeding).

\begin{figure}[t]
\begin{center}
 \includegraphics[width=4.5in]{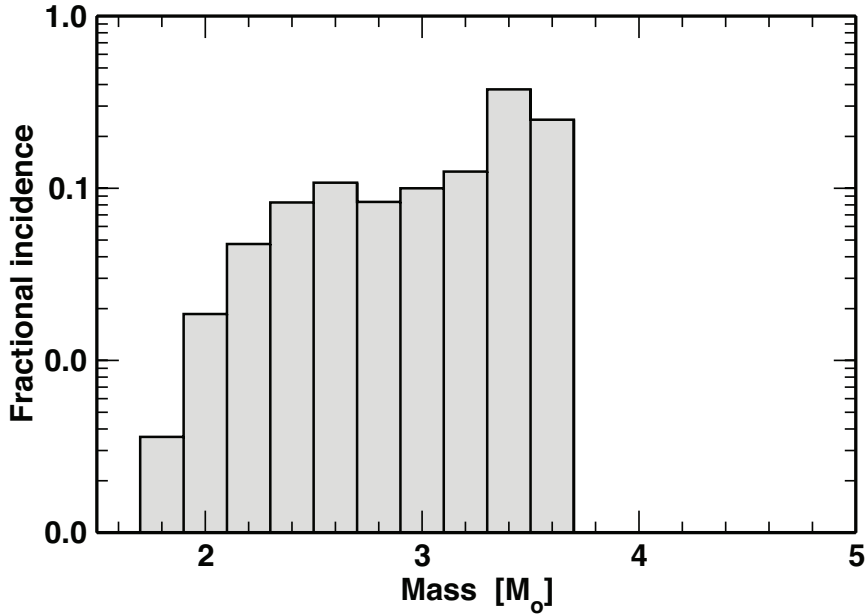}  
 \caption{Inferred incidence of magnetic fields versus mass for all intermediate-mass stars within 100 pc of the sun. A trend of increasing incidence with stellar mass is seen. The number of stars per bin decreases with mass, due to the IMF. For example, the 2\,M$_\odot$ bin contains a total of 438 stars (8 magnetic stars) whereas the 3.6\,M$_\odot$ bin contains 11 stars (3 magnetic stars). From Power (2007).  }
\label{fig_power}
\end{center}
\end{figure}

Historically, it has been assumed that magnetic fields in OB stars are very rare, and perhaps 
altogether absent in stars with masses above 8 solar masses. However, the increasing discoveries of fields in early B-type stars on the main sequence and pre-main sequence and in both young and evolved O-type stars show convincingly that fossil fields can and do exist in stars with masses as large as 45 solar masses. Given that the detected fields are sufficiently weak (0.3-1.5 kG) to have remained undetected by previous generations of instrumentation, and that recent observational results suggest that the fraction of magnetic stars increases toward higher masses (see Fig. \ref{fig_power}; Power et al. 2007), it may well be that magnetic fields are far more common in OB stars than has been supposed. Some preliminary studies of incidence in OB clusters have been performed, although with limited samples. The incidence of magnetic fields in OB stars seems indeed widespread, from a tenth to a third of the massive star population (Petit et al. 2008, this proceeding; McSwain 2008). Large scale spectropolarimetric surveys - like the Magnetism in Massive Stars Large Program (Wade et al. this proceeding) - will provide a sample large enough to derive more precise incidences.

The magnetic fields of ApBp stars are closely tied with abundance anomalies at their surface. 
In fact, it has been shown that all firmly classified Ap/Bp stars show detectable surface magnetic fields (Auri\`ere et al. 2007). The first magnetic hot OB stars to be discovered were main 
sequence helium-weak and helium-strong stars (Borra \& Landstreet 1979, Borra, Landstreet \& Thompson 1983), objects which also display strong photospheric chemical abundance anomalies, thought to be the extension of the ApBp phenomenon to higher temperatures.
However, magnetic fields are also found in chemically normal B-type stars of surface temperature similar to He-strong stars, a coexistence that does not occur for ApBp stars.  Furthermore, magnetic fields are found in a broad range of high-mass objects; emision-line stars, pulsating stars, X-ray bright stars, etc.

Although some magnetic detections have been reported for classical Be stars, none have been currently confirmed. Therefore, large-scale magnetic fields do not seem related to the Be phenomenon, nor the Oe phenomenon (e.g. Fullerton et al. this proceeding). This is not surprising considering the theoretical difficulties in a Keplerian rotation profile from a magnetically-supported disc (Owocki 2004). However, a rigid rotation disk can be produced by a large scale magnetic field. The prototype for stars with such discs is the magnetic B-type star $\sigma$\,Ori\,E (see Oksala et al. this proceeding), for which the peculiar photometric variation, the \textit{periodic} variation of its H$\alpha$ line profile, and its UV and X-ray variability have been coherently  explained by a strongly magnetically confined wind, co-rotating with the star (Townsend et al. 2005). This scenario is strengthened by the recent discovery of similar stars, for example HR\,7355 (Oksala et al. 2010, this proceeding; Rivinius et al. 2010, this proceeding) and HD\,142184 (Grunhut et al. this proceeding). 

A large incidence of magnetic stars in slowly pulsating B-type stars and $\beta$\,Cephei-like stars was reported by Hubrig et al. (2006, 2009). However, according to Silvester et al. (2009), only 1 SPB star ($\xi^1$\,CMa, see Fourtune-Ravard et al. this proceeding) and 1 $\beta$\,Cep star (16\,Peg) were confirmed in a sub-sample (containing 12 SPB stars and 7 $\beta$\,Cep-like stars) of Hubrig et al. sample - where 8 stars were claimed detected. Outside of the study of Silvester et al, there are only a few cases of confirmed magnetic fields in $\beta$\,Cephei and SPB stars, namely $\beta$ Cep itself (Henrichs et al. 2000), V2052\,Oph (Neiner et al. 2003b) and $\zeta$\,Cas (Neiner et al. 2003a).

Magnetic fields are also found in hot stars presenting periodic variability in their UV wind lines, for example $\sigma$\,Lupi (Henrichs et al. this proceeding) and $\tau$\,Sco (Donati et al. 2006b; Petit et al. this proceeding).

Confirmed magnetic O-type stars are still rare. To date, repeated detections of circular polarization within line profiles have firmly established the presence of magnetic fields 
in 5 O-type stars ($\theta^1$\,Ori\,C, HD\,191612, $\zeta$\,Ori\,A, HD\,57682 and HD\,108; Donati et al. 2002, 2006a; Bouret et al. 2008; Grunhut et al. 2009; Martins et al. 2010). These O-type stars show a wide variety of properties. However, some common characteristics can be tentatively identified. All these stars present periodic variation of H$\alpha$.
Most of these stars present anomalously slow rotation compared to the bulk of the O star population. HD\,191612 and HD\,108 are member of the Of?p class - which displays strong C\textsc{iii}\,4650 emission lines and narrow P\,Cygni/emission features (see Naz\'e et al. 2008, this proceeding). In addition to HD\,108 and HD\,191612, a magnetic detection has been reported in the third prototypical Of?p star - HD\,148937 - by Wade et al. (this proceeding).


\begin{thebibliography}{}

\bibitem[]{Auriere07}
{Auri\`ere}, M., {Wade}, G.~A., {Silvester}, J., {Ligni\`eres}, F., {Bagnulo}, S., et al. 2007, \textit{A\&A} 475, 1053

\bibitem[Bagnulo et al. 2002]{Bagnulo02}
{Bagnulo}, S., {Szeifert}, T., {Wade}, G.~A., {Landstreet}, J.~D. \& {Mathys}, G 2002, \textit{A\&A} 389, 191

\bibitem[Bagnulo et al. 2004]{Bagnulo04}
{Bagnulo}, S., {Hensberge}, H., {Landstreet}, J.~D., {Szeifert}, T. \& {Wade}, G.~A. 2004 \textit{A\&A} 416, 1149

\bibitem[]{Borra79}
{Borra}, E.~F. \& {Landstreet}, J.~D. 1979, \textit{ApJ} 228, 809

\bibitem[]{Borra83}
{Borra}, E.~F., {Landstreet}, J.~D. \& {Thompson}, I. 1983, \textit{ApJS} 53, 151

\bibitem[]{Bouret08}
{Bouret}, J.-C., {Donati}, J.-F., {Martins}, F., {Escolano}, C. et al. 2008, \textit{MNRAS} 398, 75

\bibitem[]{Braithwaite09}
{Braithwaite}, J. 2009, \textit{MNRAS} 397, 763

\bibitem[]{grunhut09}
{Grunhut}, J.~H., {Wade}, G.~A., {Marcolino}, W.~L.~F., {Petit}, V. et al. 2009, \textit{MNRAS} 400, L94

\bibitem[]{Hartmann1987}
{Hartmann}, L.~W. \& {Noyes}, R.~W. 1987, \textit{ARA\&A} 25, 271

\bibitem[]{Donati02} 
{Donati}, J.-F., {Babel}, J., {Harries}, T.~J., {Howarth}, I.~D. et al. 2002, \textit{MNRAS} 333, 55

\bibitem[]{Donati06a} 
{Donati}, J.-F., {Howarth}, I.~D., {Bouret}, J.-C., {Petit}, P. et al. 2006a, \textit{MNRAS} 365, L6

\bibitem[]{donati06} 
{Donati}, J.-F., {Howarth}, I.~D., {Jardine}, M.~M., {Petit}, P. et al. 2006b, \textit{MNRAS} 370, 629

\bibitem[Donati et al. (1997)]{Donati97}
{{Donati}, J.-F., {Semel}, M., {Carter}, B., {Rees}, D. \& {Collier Cameron}, A. 1997,
\textit{MNRAS}, 291, 658}

\bibitem[]{duez10}
{Duez}, V. \& {Mathis}, S. 2010, \textit{A\&A} 517, A58

\bibitem[]{henrichs00}
{Henrichs}, H.~F. et al. 2000, in ASP Conf. Ser. Vol. 214, IAU Colloq. 175, p. 324 

\bibitem[Hillier \& Miller (1998)]{Hillier98}
{Hillier}, D.~J. \& {Miller}, D.~L 1997, \textit{ApJ} 496, 407

\bibitem[]{Hubrig06}
{Hubrig}, S., {Briquet}, M., {Schöller}, M., {De Cat}, P. et al. 2006, \textit{MNRAS} 369, L61

\bibitem[]{Hubrig09}
{Hubrig}, S., {Briquet}, M., {De Cat}, P., {Schöller}, M., {Morel}, T. \& {Ilyin}, I. 2009, \textit{AN} 330, 317

\bibitem[]{Kochukhov06}
{Kochukhov}, O. \& {Bagnulo}, S. 2006, \textit{A\&A} 450, 763

\bibitem[]{Kochukhov10b}
{Kochukhov}, O., {Makaganiuk}, V. \& {Piskunov}, N. 2010, \textit{A\&A}, accepted

\bibitem[Kochukhov \& Wade (2010)]{Kochukhov10}
{Kochukhov}, O. \& {Wade}, G.~A. 2010, \textit{A\&A} 513, A13

\bibitem[Landstreet (1982)]{Landstreet82}
{Landstreet}, J.~D. 1982, \textit{ApJ} 258, 639

\bibitem[Landstreet et al. (2007)]{Landstreet07}
{Landstreet}, J.~D., {Bagnulo}, S., {Andretta}, V., {Fossati}, L. et al. 2007, \textit{A\&A} 470, 685

\bibitem[Landstreet \& Mathys (2000)]{Landstreet00}
{Landstreet}, J.~D. \& {Mathys}, G. 2000, \textit{A\&A} 359, 213

\bibitem[Maeder \& Meynet (2005)]{Maeder05}
{Maeder}, A. \& {Meynet}, G. 2005, \textit{A\&A} 440, 1041

\bibitem[]{mcswain08}
{McSwain}, M.~V. 2008, \textit{ApJ} 686, 1269

\bibitem[]{Naze08}
{Naz\'e}, Y., {Walborn}, N.~R. \& {Martins}, F. 2008, \textit{Rev. Mexicana AyA} 44, 331

\bibitem[]{Oksala10}
{Oksala}, M.~E., {Wade}, G.~A., {Marcolino}, W.~L.~F., {Grunhut}, J. et al. 2010, \textit{MNRAS} 405, L51

\bibitem[]{Owocki04}
{Owocki}, S.~P. 2004, IAU Symposium No. 215, p.\ 515

\bibitem[]{Neiner03a} 
{Neiner}, C., {Geers}, V.~C., {Henrichs}, H.~F., {Floquet}, M., {Frémat}, Y. et al. 2003a, \textit{A\&A} 406, 1019

\bibitem[]{Neiner03b} 
{Neiner}, C., {Henrichs}, H.~F., {Floquet}, M., {Fr\'emat}, Y., {Preuss}, O. et al. 2003b, \textit{A\&A} 411, 565

\bibitem[Petit et al. (2008)]{Petit08}
{Petit}, V.,{Wade}, G.~A., {Drissen}, L.,{Montmerle}, T. \& {Alecian}, E. 2008, \textit{MNRAS} 387, L23

\bibitem[Piskunov \& Kochukhov 2002]{Piskunov02}
{Piskunov}, N. \& {Kochukhov}, O. 2002, \textit{A\&A} 381, 736

\bibitem[]{Power07}
{Power}, J. 2007, MSc. thesis, Queen's University (Canada)

\bibitem[Rivinius et al. (2010)]{Rivi10}
{Rivinius}, T., {Szeifert}, T., {Barrera}, L., {Townsend}, R.~H.~D. et al. 2010, \textit{MNRAS} 405, L46

\bibitem[Silvester et al. (2009)]{Silvester09}
{Silvester}, J., {Neiner}, C., {Henrichs}, H.~F., {Wade}, G.~A., {Petit}, V. et al. 2009, \textit{MNRAS} 398, 1505

\bibitem[]{Townsend05}
{Townsend}, R.~H.~D., {Owocki}, S.~P. \& {Groote}, D. 2005, \textit{ApJL} 630, L81

\bibitem[]{udDoula09}
{ud-Doula}, A., {Owocki}, S.~P., {Townsend}, R.~H.~D. 2009, \textit{MNRAS} 392, 1022

\bibitem[]{wade00}
{Wade}, G.~A., {Donati}, J.-F., {Landstreet}, J.~D. \& {Shorlin}, S.~L.~S. 2000, \textit{MNRAS} 313, 823





\end{thebibliography}
\end{document}